\documentclass[conference]{IEEEtran}
\pdfoutput=1

\usepackage[utf8x]{inputenc}
\usepackage{amsmath}
\usepackage{xargs}
\usepackage{todonotes}
\newcommandx{\todoLukas}[2][1=]{\todo[linecolor=red,inline,backgroundcolor=red!25,bordercolor=red,#1]{\textbf{Lukas: }#2}}
\newcommandx{\todoNishant}[2][1=]{\todo[linecolor=blue,inline,backgroundcolor=blue!25,bordercolor=blue,#1]{\textbf{Nishant: }#2}}

\usepackage{subcaption}
\usepackage{environ}
\makeatletter
\newsavebox{\measure@tikzpicture}
\NewEnviron{scaletikzpicturetowidth}[1]{%
  \def\tikz@width{#1}%
  \def\tikzscale{1}\begin{lrbox}{\measure@tikzpicture}%
  \BODY
  \end{lrbox}%
  \pgfmathparse{#1/\wd\measure@tikzpicture}%
  \edef\tikzscale{\pgfmathresult}%
  \BODY
}
\makeatother

\usepackage{csquotes}
\usepackage[nolist]{acronym}
\usepackage{hyperref}
\usepackage[inline,shortlabels]{enumitem}
\usepackage{cleveref}

\usepackage{xcolor}
\usepackage{pgfplots}
\pgfplotsset{compat=newest}
\usepackage{pgfplotstable}
\usepackage{pgf-umlsd}
\usepackage{colortbl}
\colorlet{negro}{black}
\colorlet{gris}{black!70}
\colorlet{rojo}{red!70!black}
\colorlet{rojol}{red}
\usepackage{booktabs}
\usepackage{tikz}
\usetikzlibrary{positioning}
\usetikzlibrary{petri} 
\usetikzlibrary{backgrounds}
\usetikzlibrary{calc,shapes.arrows}
\usepackage{pgfplotstable}
\usepackage{pgf-umlsd}
\usepackage{flushend}
\usepackage{tabularx}
\usepackage{color} 

\begin{document}

\title{Attack-aware Security Function Chain Reordering}

\makeatletter
\newcommand{\linebreakand}{%
  \end{@IEEEauthorhalign}
  \hfill\mbox{}\par
  \mbox{}\hfill\begin{@IEEEauthorhalign}
}
\makeatother

\author{\IEEEauthorblockN{Lukas Iffländer, Lukas Beierlieb, Nicolas Fella, Samuel Kounev}
\IEEEauthorblockA{University of Würzburg\\
Germany\\
firstname.lastname@uni-wuerzburg.de}
\and
\IEEEauthorblockN{Nishant Rawtani, Klaus-Dieter Lange}
\IEEEauthorblockA{Hewlett Packard Enterprise\\
India/USA\\
nishant.rawtani/klaus.lange@hpe.com}}

\maketitle
\thispagestyle{plain}
\pagestyle{plain}

\IEEEpeerreviewmaketitle

\begin{acronym}[WS-Agreement] 
    \acro{ACK}{TCP ACK (Acknowledgment) packet}
    \acro{API}{Application Programming Interface}
    \acro{ARP}{Address Resolution Protocol}
    \acro{ASIC}{Application Specific Integrated Circuit}
    \acro{AVX}{Advanced Vector Extensions}
    \acro{CPU}{Central Processing Unit}
    \acro{CSA}{Cloud Security Alliance}
    \acro{CVE}{Common Vulnerabilities and Exposures}
    \acroplural{CVEs}{Common Vulnerabilities and Exposures}
    \acro{DMZ}{Demilitarized Zone}
    \acro{DPDK}{Data Plane Development Kit}
    \acro{DDoS}{Distributed Denial-of-Service}
    \acro{DoS}{Denial-of-Service}
    \acro{DPI}{Deep Package Inspection}
    \acro{DPS}{DDoS Protection System}
    \acro{EAL}{Environment Abstraction Layer}
    \acro{FCC}{Function Chaining Controller}
    \acro{FCFS}{First Come First Serve}
    \acro{FPGA}{Field-Programmable Gate Array}
    \acro{FTP}{File Transfer Protocol}
    \acro{GET}{HTTP request to require data from an HTTP server}
    \acro{HTTP}{Hypertext Transfer Protocol}
    \acro{IaaS}{Infrastructure as a Service}
    \acro{ICMP}{Internet Control Message Protocol}
    \acro{IDPS}{Intrusion Detection and Prevention System}
    \acroplural{IDPS}[IDPSes]{Intrusion Detection and Prevention Systems}
    \acro{IDS}{Intrusion Detection System}
    \acroplural{IDS}[IDSes]{Intrusion Detection Systems}
    \acro{IoT}{Internet of Things}
    \acro{IP}{Internet Protocol}
    \acro{IPS}{Intrusion Prevention System}
    \acroplural{IPS}[IPSes]{Intrusion Prevention Systems}
    \acro{ISN}{Initial Sequence Number}
    \acro{I/O}{Input/Output}
    \acro{JSON}{JavaScript Object Notation}
    \acro{JWE}{\acs{JSON} Web Encryption}
    \acro{JWS}{\acs{JSON} Web Signature}
    \acro{JWT}{\acs{JSON} Web Token}
    \acro{KVM}{Kernel Virtual Machine}
    \acro{L2}{Layer 2}
    \acro{LAN}{Local Area Network}
    \acro{MAC}{Media Access Control}
    \acro{MAC*}{Message Authentication Code}
    \acro{MSS}{Maximum Segment Size}
    \acro{NAT}{Network Address Translation}
    \acro{NFV}{Network Function Virtualization}
    \acro{NIC}{Network Interface Card}
    \acro{NIDS}{Network Intrusion Detection System}
    \acroplural{NIDS}[NIDS]{Network Intrusion Detection Systems}
    \acro{NMS}{Network Management System}
    \acro{ODL}{OpenDaylight}
    \acro{ONOS}{Open Network Operating System}
    \acro{OF}{OpenFlow}
    \acroindefinite{OF}{an}{an}
    \acro{OFA}{OpenFlow Agent}
    \acro{OID}{Object Identifier}
    \acro{OS}{Operating System}
    \acro{OSI}{Open Systems Interconnection}
    \acro{PMD}{Poll Mode Driver}
    \acro{PoC}{proof-of-concept}
    \acro{POST}{HTTP request to submit data so an HTTP server (e.g., via a form)}
    \acro{QoE}{Quality of Experience}
    \acro{QoS}{Quality of Service}
    \acro{REST}{Representational State Transfer}
    \acro{SDN}{Software-defined Networking}
    \acro{SECaaS}{Security as a Service}
    \acro{SFC}{Service Function Chain}
    \acro{SSFC}{Security Service Function Chain}
    \acro{SFCing}{Security Function Chaining}
    \acro{SSFCing}{Security Service Function Chaining}
    \acro{SFCC}{Security Function Chaining Controller}
    \acro{SIMD}{Single Instruction Multiple Input}
    \acro{SNMP}{Simple Network Management Protocol}
    \acro{SR-IOV}{Single-Root I/O Virtualization}
    \acro{SSE}{Streaming \ac{SIMD} Extensions}
    \acro{SSH}{Secure Shell}
    \acro{SYN}{TCP SYN (Synchronization) packet}
    \acro{SYN+ACK}{TCP SYN+ACK (Synchronization and Acknowledgment) Packet}
    \acro{TCB}{Transmission Control Block}
    \acro{TCP}{Transmission Control Protocol}
    \acro{THREADS}{TCP Handshake Remote Establishment and Dynamic Rerouting using SDN}
    \acro{TLS}{Transport Layer Security}
    \acro{UDP}{User Datagram Protocol}
    \acro{URI}{Uniform Resource Identifier}
    \acro{URL}{Uniform Resource Locator}
    \acro{URG}{TCP Urgent Flag}
    \acro{VLAN}{Virtualized Local Area Network}
    \acro{VM}{Virtual Machine}
    \acro{VNF}{Virtualized Network Function}
    \acro{WAF}{Web Application Firewall}
    \acro{WSGI}{Web Server Gateway Interface}
\end{acronym}

\begin{abstract}
    Attack-awareness recognizes self-awareness for security systems regarding the occurring attacks. 
    More frequent and intense attacks on cloud and network infrastructures are pushing security systems to the limit.
    With the end of Moore's Law, merely scaling against these attacks is no longer economically justified.
    Previous works have already dealt with the adoption of \acl{SDN} and \acl{NFV} in security systems and used both approaches to optimize performance by the intelligent placement of security functions.
    However, these works have not yet considered the sequence in which traffic passes through these functions.
    In this work, we make a case for the need to take this ordering into account by showing its impact.
    We then propose a reordering framework and analyze what aspects are necessary for modeling security service function chains and making decisions regarding the order based on those models.
    We show the impact of the order and validate our framework in an evaluation environment.
    The effect can extend to multiple orders of magnitude, and the framework's evaluation proves the feasibility of our concept.
\end{abstract}

\section{Introduction}

Today's network attacks rely on massive bot networks. Their attacking power rises as the number of online devices rapidly grows in times of \ac{IoT}.
The ending of Moore's Law  (promising doubled resources every two years) limits the opportunity to throw in additional resources to fight attacks.
Moreover, booking additional resources on demand is very costly, especially considering that the owners of bot networks do not have to pay for their attack resources. 

IT systems providing services via a network offer various attack vectors.
For each type of network attack, there are dedicated security functions to defend the system.
Multiple security functions together form \acp{SSFC} to protect a system against a set of attack types.
For most systems, there is a direct correlation between consumed resources and the number of processed packages.
In contrast, security functions (and therefore \acp{SSFC}) stand out, as they drop packets deemed as malicious causing lower load on subsequent security functions.

\begin{figure}[!b]
    \centering
    \includegraphics[width=\columnwidth]{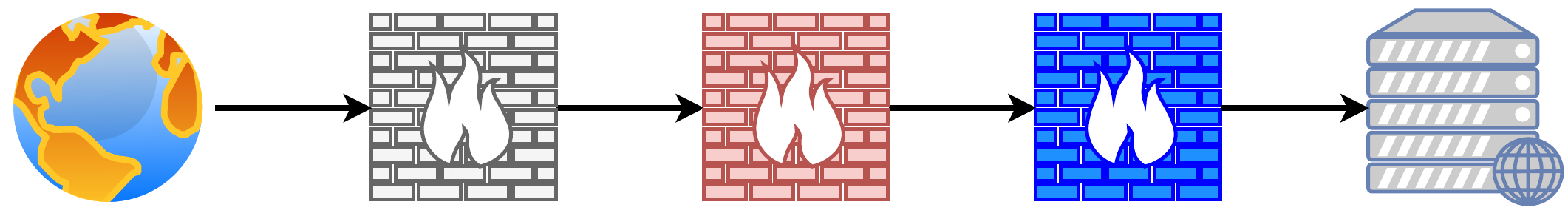}
    \caption{An example for a \acl{SSFC}.}
    \label{fig:dynamic-reordering:general-idea:initial}
\end{figure}

\Cref{fig:dynamic-reordering:general-idea:initial} shows an \ac{SSFC} with three different security functions. 
In most current security architectures, those \acp{SSFC} are hard-wired or interconnected using a fixed order via \ac{SDN}.

If the setup is under standard load without an attack occurring, the \ac{SSFC} order is not relevant. 
All packets are benign and, therefore, have to pass through all security functions in the \ac{SSFC}.
Reordering the functions would not change the resource demand generated by the benign packets.

This changes when an attack occurs, that matches one of the security function.
With the initial configuration, the traffic passes through all security functions in the \ac{SSFC}. 
Thus, it creates resource demand at every step until the last security function stops it.
When we put the blue security function at the front of the \ac{SSFC},
this order leads to the first security function dropping the malicious traffic immediately. 
Thus, traffic does not pass on to the white and red security functions and creates only resource demands at the blue security function.
\Cref{tab:dynamic-reordering:general-idea:example-performance} shows an example scenario which proves that optimal configurations can reduce the total number of required instances.

\begin{table}[!b]
    \centering
    \begin{tabular}{lcccc}
        \toprule
        SSFC & \multicolumn{4}{c}{number of instances}\\
        Ordering & red & white & blue & total \\
         \midrule
         red -- white -- blue & 10 & 5 & 7 & 22\\
         red -- blue -- white & 10 & 7 & 1 & 18\\
         white -- red -- blue & 5 & 10 & 7 & 22\\
         white -- blue -- red & 5 & 7 & 1 & 13\\
         blue -- red -- white & 7 & 1 & 1 & 9\\
         blue -- white -- red & 7 & 1 & 1 & 9\\
         \bottomrule
    \end{tabular}
    \caption{Example calculation of the resource demand for different \ac{SSFC} orders of the example \ac{SSFC}. Throughput per instance: red 100~MBit/s, blue 150~MBit/s, white 200~MBits/s. Load profile: 950~MBit/s malicious traffic matching the blue security function and 50~MBit/s of benign traffic.}
    \label{tab:dynamic-reordering:general-idea:example-performance}
\end{table}

Related work mainly deals with \ac{NFV} as an enabler for security functions covered \ac{SDN} either a security risk or feature.
To the best of our knowledge, no other works went into analyzing the performance impact of \ac{SSFC} orders or presented solutions and models for \ac{SSFC} reordering.

This work introduces the following \emph{contributions}:
\begin{itemize}
    \item An analysis of the performance impact of different individual security functions and of security functions in \acp{SSFC} with different orders.
    \item A design, \ac{PoC} implementation, and evaluation of a framework for dynamic \ac{SSFC} reordering.
    \item Modeling formalisms for the traffic inside the network, single security functions and \acp{SSFC}.
\end{itemize}

The remainder of this paper is structured as follows:
First, \Cref{sec:foundations} and \Cref{sec:related} present the required background and related works.
We analyze security function and \ac{SSFC} performance in \Cref{sec:impact} and present our \ac{SSFC} reordering framework in \Cref{sec:framework}.
\Cref{sec:dynamic-reordering:modeling} introduces modeling approaches for individual security functions and \acp{SSFC}, and \Cref{sec:conclusion} provides our conclusion and future work.
\section{Foundations}
\label{sec:foundations}

For our approach, we use various security functions.
Also, \ac{SDN} and \ac{NFV} are important underlying technologies.

\subsection{Security Appliances}

We use multiple security functions in this work.
We mainly focus on \acsp{DPS}, \acsp{IDPS}, and firewalls.

\subsubsection{Distributed Denial-of-Service (DDoS) Protection Systems}
\acused{DPS}

One of the most popular mechanism to mitigate \acs{DDoS} attacks is SYN Cookies.
It is readily available for services running on top of mainline Linux kernel, and therefore is widely adopted.
A SYN flooding attack exploits the limited size of the \acs{TCP} buffer which is a critical resource for establishing new connections.
SYN cookies
are a \ac{TCP} standard compliant way of eliminating the need for buffer entries related to half-open connections.
Generally, the data stored in the buffer is necessary to check if a received ACK packet belongs to previous SYN and SYN+ACK packets and whether the client received the server’s initial sequence number correctly.
The idea of SYN cookies is to not store this information locally, but to encode it into the sent SYN+ACK packet and to retrieve this information from the ACK response.

\subsubsection{\acfp{IDPS}}
\acp{IDPS} combine \acp{IDS} and \acp{IPS}.
\acsp{IDS} can detect attacks
and provide additional defense mechanisms, whereas, \acsp{IPS} are capable of actively defending against incoming attacks.
\acp{IDPS} can be classified based on the type of Monitored Platform, Attack Detection Methods used, Monitoring Method and Deployment Architecture.
In this work, we focus on network-based, misused-based, real-time and non-distributed \acp{IDPS}.
A network-based \acp{IDPS} is placed strategically on the network to detect any attacks that originate from outside the network.
Misuse-based \acp{IDPS} primarily target singular attacks that usually are carried out in a single step~\cite{Vigna_2004_TestingNetwork-basedIntrusionDetectionSignaturesUsingMutantExploits} to exploit a selected vulnerability.
Here, an \acp{IDPS} uses signatures containing the features of the exploit for its detection.
Real-time \acp{IDPS} intercept the packets before they reach the target system and work synchronously to the traffic flow.
Non-distributed \acp{IDPS} are deployed at a singular (central) position inside the network.

\subsubsection{Firewalls}

Firewalls can be defined~\cite{Oppliger_1997_CommunicationsoftheACM_InternetSecurityFirewallsandBeyond} as an intermediate system that is plugged between the network and the Internet to establish a controlled link, and to erect an outer security wall or perimeter.
The aim of this perimeter is to protect the network from network-based threats and attacks, and to provide a single choke point where security and audit can be imposed.
Most common firewalls work on the third layer of the OSI-Stack, also known as the network layer.
These firewalls filter the incoming packets based on a pre-defined set of rules and check whether a packet matches against these rules or not.
The rules rely on the information available in the packet headers such as protocol numbers, source and destination IP addresses.
Another type of firewall is the so-called proxy servers (e.g., SYNPROXY).
These servers require authentication before the individual services can be accessed.
If the authentication is successful, the proxy forwards packets between the server and the client.

\subsection{\acf{SDN}}

\ac{SDN} takes on the challenges posed by the increasing number of participants in the network and the associated exponential rise in cost due to the directly correlated growth in resource demands.
The objective of \ac{SDN} was to achieve greater scalability, flexibility, automation, and independence from hardware manufacturers to reduce acquisition and operating costs.
\ac{SDN} relies on four basic principles:

The \emph{separation of control and data planes} divides the switching process into the control plane, using routing algorithms to decide on packet forwarding, with the data plane, technically handling the packet.
\ac{SDN} allows influencing the forwarding process externally to communicate with the switch allowing the switch to change its behavior at runtime without having to replace the hardware components.

The \emph{central control instance}, called the controller, enables the configuration and administration of the network.
For reasons of availability or load distribution, it can be deployed as a physical or virtual replica.

The behavior of a switch can be changed using software, enabling the installation from algorithms or other applications from different manufacturers - independent of the hardware producer.
This feature also allows applications to operate above the network layer, i.e., at the application level, regardless of the switch model or the \acs{OS}.

SDN brings together areas that are traditional handled separately via four essential \acsp{API}.
The \emph{Southbound \acs{API}} connects the control and data layer.
\acl{OF} is the most popular open protocol for the Southbound \acs{API} in \acp{SDN}.
\ac{OF} can be used to configure and evaluate the statistics of a network device, usually a switch.
The \emph{Westbound \acs{API}} is used to communicate different control layers of different domains.
The \emph{Northbound \acs{API}} exchanges information between the application and control layers.
The \emph{Eastbound \acs{API}} provides a contact surface for non-SDN components.

\subsection{\acf{NFV}}

\ac{NFV} is a new paradigm for networks.
Typically deployed on proprietary specialized hardware in the past, these functions are replaceable by software solutions running on commodity hardware~\cite{Chiosi_2012_NetworkFunctionsVirtualizationNFVanIntroductionBenefitsEnablersChallenges&CallforAction,Rygielski_2017_FlexibleModelingofDataCenterNetworksforCapacityManagement}.
Typical examples of such functions are switching, routing, load balancing, and firewalls.
The implementation of a function usually is referenced as a \ac{VNF} as it is commonly deployed inside a \ac{VM} to allow for higher flexibility and scalability.
Not every function is suitable for conversion into a \ac{VNF}.
The use of multiple resources having optimized processors or \acsp{FPGA} still can be advantageous for real-time requirements. 
\acp{VNF} depend on performance in several ways.
First, the network adapters limit the number and the speed of available ports.
Second, the I/O subsystem between the network card and the application can affect the performance.
Also, the resources provided for the application (such as memory, and \ac{CPU}) can become a bottleneck.
Many \ac{NFV} solutions are implemented usually in conjunction with specialized \acp{OS} or drivers to minimize bottlenecks.
Mapping the network functions in the software separates the data and control layers, one of the central goals of SDN.
\section{Related Work}
\label{sec:related}

We provide a compressed overview of works regarding \ac{SDN} for security and security for \ac{SDN}. 
We further present works on \ac{NFV} deployment, application, and the whitepaper that motivated our research.

\ac{SDN} creates \enquote{a very fascinating dilemma}~\cite{Yan_2015_IEEECommunicationsMagazine_DistributedDenialofServiceAttacksinSoftware-definedNetworkingwithCloudComputing} --- it offers a wide range of benefits for security implementation while introducing new security challenges.
The research demonstrated that different attacks can affect not only the functioning of the targeted component but also the availability and confidentiality of each layer and interface of the \ac{SDN} stack.
Scott-Hayward et al.~\cite{Scott-Hayward_2016_IEEECommunicationsSurveys&Tutorials_ASurveyofSecurityinSoftwareDefinedNetworks} presented a detailed analysis of these security challenges and categorized them by the affected \ac{SDN} layer/interface.

In addition to traditional \emph{\ac{DoS}} attacks, the intelligence centralization and vertical split into three main functional layers that expand the attack surface and inspire new techniques for each layer.
An \emph{Application Layer \ac{DoS} Attack} directly targets an application to consume all of the resources allocated to it and to cause a \ac{DoS}.
A \emph{Control Layer \ac{DoS} Attack} can arise by targeting any of its components, (e.g., forcing different applications to generate many conflicting flow rules may lead a controller to an unpredictable state).
In \emph{Infrastructure Layer \ac{DoS} Attacks}, bottlenecks in \ac{OF} switches and the southbound \ac{API} are exploited.
Moreover, generating fake flows fills the flow tables and prevents rules for normal network flows to be stored~\cite{Sezer_2013_IEEECommunicationsMagazine_ArewereadyforSDN?Implementationchallengesforsoftware-definednetworks}.

\ac{SDN} provides opportunities to revisit old security concepts and to introduce new techniques as \ac{SDN} features~\cite{Yan_2015_IEEECommunicationsMagazine_DistributedDenialofServiceAttacksinSoftware-definedNetworkingwithCloudComputing} to enhance network resilience.
Network-wide knowledge facilitates the validation of security policies and enables quick identification and resolution of any conflicts~\cite{McBride_2013_OpenNetworkingFoundation-ONFSOLUTIONBRIEF_SDNsecurityconsiderationsinthedatacenter}.
As a result, consistent security policies can be built and maintained.
In addition, \ac{SDN} supports software-based traffic analysis that opens the door for innovative ideas, and Chi et al.~\cite{Chi_2014_AnAMIThreatDetectionMechanismBasedonSDNNetworks} present different concepts on how to integrate the Snort \ac{IDS} into an \ac{SDN}-based network.

Barbette et al.~\cite{Barbette_2015_FastUserspacePacketProcessing}, as well as Gallemnueller et al.~\cite{Gallenmuller_2015_ComparisonofFrameworksforHigh-performancePacketIO}, provide a detailed comparison of various \ac{NFV} development kits and frameworks, including the recently proposed XDP framework~\cite{Hoiland-Jorgensen_2018_TheeXpressDataPathFastProgrammablePacketProcessingintheOperatingSystemKernel}.
In the context of the performance evaluation of software-implemented network functions, \cite{Chiosi_2012_NetworkFunctionsVirtualizationNFVanIntroductionBenefitsEnablersChallenges&CallforAction} provides an extensive list of best practices and caveats.
A detailed network security investigation related to the use cases of \ac{SDN} and \ac{NFV} is presented in the survey by Lorenz et al.~\cite{Lorenz_2017_IEEECommunicationsMagazine_AnSDN/NFV-EnabledEnterpriseNetworkArchitectureOfferingFine-GrainedSecurityPolicyEnforcement}, and Farris et al. provide an extensive overview of emerging \ac{SDN} and \ac{NFV} security mechanisms in the context of the Internet of Things~\cite{Farris_2019_IEEECommunicationsSurveys&Tutorials_ASurveyonEmergingSDNandNFVSecurityMechanismsforIoTSystems}.

In addition to introducing the possibilities for dynamic \ac{SSFC}, \ac{SDN} also allows augmenting them with traffic and application awareness~\cite{Li_2017_J.InternetServ.Inf.Secur._Application-awareandDynamicSecurityFunctionChainingforMobileNetworks}.
Many solutions rely on \ac{SDN} and \ac{NFV} in the context of \acs{DDoS} resiliency.
Multiple open topics are discussed by various authors that include
\begin{enumerate*}
    \item rule anomalies \cite{Li_2018_IEEEAccess_RuleAnomaly-FreeMechanismofSecurityFunctionChainingin5G}, 
    \item intelligent positioning \cite{Luizelli_2015_PiecingTogethertheNFVProvisioningPuzzleEfficientPlacementandChainingofVirtualNetworkFunctions}, and
    \item effective provisioning \cite{Shameli-Sendi_2019_IEEETransactionsonServicesComputing_EfficientProvisioningofSecurityServiceFunctionChainingUsingNetworkSecurityDefensePatterns}
\end{enumerate*}.

Security Function Chaining (SFCing) is a significant component of this work and essential for complex \ac{NFV} security frameworks.
The first inspiration for this work was the \ac{CSA} \enquote{Security Position Paper: Network Function Virtualization} paper by Milenkoski et al.~\cite{Milenkoski_2016_SecurityPositionPaperNetworkFunctionVirtualization} that proposes six \ac{NFV} security challenges:
\begin{enumerate*}
    \item Hypervisor dependencies,
    \item Elastic network boundaries,
    \item Dynamic workloads,
    \item Service insertion,
    \item Stateful versus stateless inspection, and
    \item Scalability of available resources
\end{enumerate*}.
The authors detail an enterprise-grade architecture for a \ac{NFV} security framework that reduces deployment and management resources as well as adapt the \ac{SSFC} ordering of its security appliances depending on an incoming attack.
This work inspired our efforts.

\section{Impact of Security Function Chain Ordering}
\label{sec:impact}

We claim that that the \ac{SSFC} order of security functions influences the performance of the \ac{SSFC}.
In this section, we will evaluate security functions and \acp{SSFC} with different orders to assert our claim.
At first, we present the used evaluation environment.
We then measure the performance of three different security functions in a stand-alone deployment.
The measured security functions are a firewall, a \ac{DPS}, and an \ac{IDPS}.
Then, we put these security functions in \acp{SSFC} with two service functions and vary their order.
Last, we discuss the results and the conclusions that we must consider for the reordering framework and the decision-making in the following sections.

\subsection{Evaluation Environment}

\begin{figure}[tb]
	\centering
    \includegraphics[width=\columnwidth]{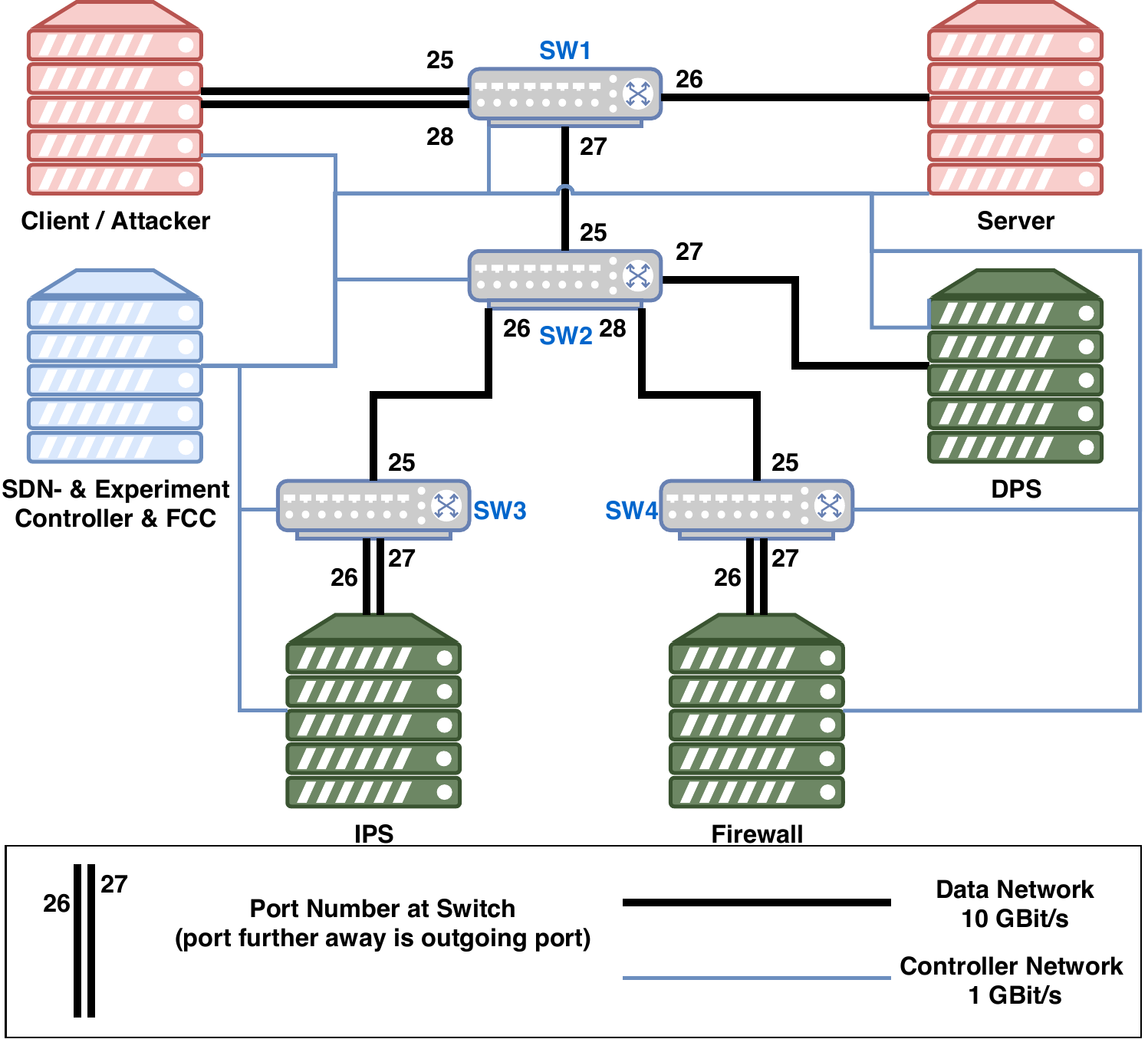}
	\caption{Evaluation setup for different security functions and \ac{SSFC} orders.}
	\label{fig:dynamic-reordering:ordering-effect:evaluation-environment:connection}
\end{figure}

To evaluate security function performance, we designed a testbed (\Cref{fig:dynamic-reordering:ordering-effect:evaluation-environment:connection}) that can incorporate benign and malicious workloads using single functions and composite \acp{SSFC} with modifiable function order and different server applications.

\subsubsection{Hardware Components}

We use a total of six physical servers.
These play the roles of \begin{enumerate*}
	\item a client and attacker,
	\item an application server (the protected application),
	\item a \acf{DPS},
	\item a firewall,
	\item an \acf{IDPS}, and 
	\item an SDN, Experiment and Function Chaining Controller
\end{enumerate*}.
For all servers, we use a four-core (8 threads) Intel Xeon E3-1230 V2 \ac{CPU} at 3.30 GHz equipped with 16 GB RAM.
The client \& attacker machine is serving two roles and, therefore, also uses one link for each role.
Two standard non-programmable 1~GBit/s HPE switches provide the connectivity for the backend and controller network and HPE 5130 24G 4SFP+ EI SDN switches span the network for the experimentation data.
The switches provide sufficient backplane switching capacity to ensure that this setup does not become a bottleneck.

\subsubsection{Software Components}
\label{sec:dynamic-reordering:ordering-effect:evaluation-environment:software-components}

\begin{enumerate*}[label=\emph{\alph*})]
    \item \emph{Traffic Generator (benign):}
        On the first 10~GBit/s interface of the client \& attacker server, we generate benign \ac{HTTP} traffic. 
        For this purpose, we use \emph{\ac{HTTP} Load Generator}~\cite{Kistowski_2018_RunTimePredictionofPowerConsumptionforComponentDeployments}.
    \item \emph{Traffic Generator (malicious):}
        We use the second 10~GBit/s interface of the client \& attacker server to create malicious packets.
        We create SYN, \ac{UDP}, and \ac{IDS} floods using Cisco's Trex generator. 
        For the chosen attacks, we use only the stateless mode.
        To generate \ac{HTTP} floods, we employ BoNeSi --- a BotNet Simulator.
    \item \emph{\ac{IDPS}:} 
        The \ac{IDPS} host runs the Snort \ac{IDPS} in version 2.9.7. 
        Snort is a popular, open-source \ac{IDPS} developed by Cisco and also is the foundation of Cisco's commercial \ac{IDPS} solutions. 
    \item \emph{Firewall:} 
        Like the \ac{IDPS}, the Firewall uses one interface for incoming and one for outgoing traffic. 
        We interconnect both interfaces using a Linux bridge, and Netfilter/iptables rules accomplish the packet filtering. 
    \item \emph{\ac{DPS}:} 
        As a \ac{DPS}, we use a modified version of \ac{THREADS}~\cite{Ifflaender_2018_AddressingShortcomingsofExistingDDoSProtectionSoftwareUsingSoftware-DefinedNetworking}. 
        \ac{THREADS} is a \ac{DPS} \ac{VNF} against SYN flood attacks.
        It handles SYN requests and only for successful requests establishes a connection with the server and triggers an \ac{SDN} reconfiguration.
        Thereby, it eliminates many shortcomings of SYNPROXY and SYN cookies.
    \item \emph{Protected Service:} 
        The target server runs TeaStore, a micro-service reference and test application emulating a basic online store~\cite{Kistowski_2018_TeaStoreAMicroServiceReferenceApplicationforBenchmarkingModelingandResourceManagementResearch}.
    \item \emph{SDN Controller:} We use Ryu as the \ac{SDN} controller. 
        The \texttt{ryu.app.ofctl\_rest} module provides a REST-based interface for deploying flows.
\end{enumerate*}

\subsubsection{Monitoring and Metrics Collection} 

The testbed measures and records the following metrics from various sources: 
\begin{enumerate*}[label=\emph{\alph*})]
	\item the \ac{CPU} usage of each server in various states: \texttt{user}, \texttt{iowait}, \texttt{softirq}, \texttt{system},
	\item the total number of sent and successful benign \ac{HTTP} requests, and
	\item the average \ac{ICMP} and \ac{TCP} SYN latency and packet loss between sender and receiver
\end{enumerate*}.
Telegraf collects \ac{CPU} usage statistics and sends them to an InfluxDB running on the experiment controller.
We use Grafana to visualize the gathered data. 
The \ac{HTTP} traffic generator reports the number of total and successful requests during the run. 
The ping command allows measuring the latency and packet loss between the sender and receiver. 
We measure the SYN latency and packet loss using hping3. 
For \ac{ICMP} and SYN latency and packet loss, we run an attack of intensity \(x\) for a time \(t\) and ping and establish \ac{TCP} connections during that time.

\subsection{Single Security Function Performance}
\label{sec:dynamic-reordering:ordering-effect:single-application-performance}

\begin{figure}[!tb]
    \begin{tikzpicture}
    \begin{axis}[
        width=.9\columnwidth,
        height=.25\textheight,
        scaled ticks=false, 
        tick label style={/pgf/number format/fixed},
        ymajorgrids = true,
        ylabel= {Successful Requests},
        ymin = -500, ymax = 1550000,
        xlabel = {Requests per Second},
        xmin = -500, xmax = 22500,
        legend style=
        {
            column sep = 10pt,
            nodes=
            {
                inner sep=2, 
                below=-1.1ex
            }, 
            at={(0.95,0.25)}, 
            anchor=south east
        }, 
    ]
        \addplot+[
            mark options={scale=.8},
            error bars/.cd, y dir=both, y explicit,
            error mark options = {
                rotate = 90, 
                mark size = 4pt, 
                gris,
            }
        ]
        table[x=load, y=perf, y error = error, col sep=comma]    {data/05_dynamic-reordering/ordering-effect/single-application-performance_direct.csv};
        \addlegendentry{direct}
        \addplot+[
            mark options={scale=.8},
            error bars/.cd, y dir=both, y explicit,
            error mark options = {
                rotate = 90, 
                mark size = 4pt, 
                gris,
            }
        ]
        table[x=load, y=perf, y error = error, col sep=comma]    {data/05_dynamic-reordering/ordering-effect/single-application-performance_firewall.csv};
        \addlegendentry{FW}
        \addplot+[
            mark options={scale=.8},
            error bars/.cd, y dir=both, y explicit,
            error mark options = {
                rotate = 90, 
                mark size = 4pt, 
                gris,
            }
        ]
        table[x=load, y=perf, y error = error, col sep=comma]    {data/05_dynamic-reordering/ordering-effect/single-application-performance_dps.csv};
        \addlegendentry{DPS}
        \addplot+[
            mark options={scale=.8},
            error bars/.cd, y dir=both, y explicit,
            error mark options = {
                rotate = 90, 
                mark size = 4pt, 
                gris,
            }
        ]
        table[x=load, y=perf, y error = error, col sep=comma]    {data/05_dynamic-reordering/ordering-effect/single-application-performance_ids.csv};
        \addlegendentry{IDPS}
    \end{axis}
\end{tikzpicture}
    \caption{Performance (in successful requests over 90 seconds) for the Direct Chain and single intermediary security functions.}
    \label{fig:benign:overview}
\end{figure}
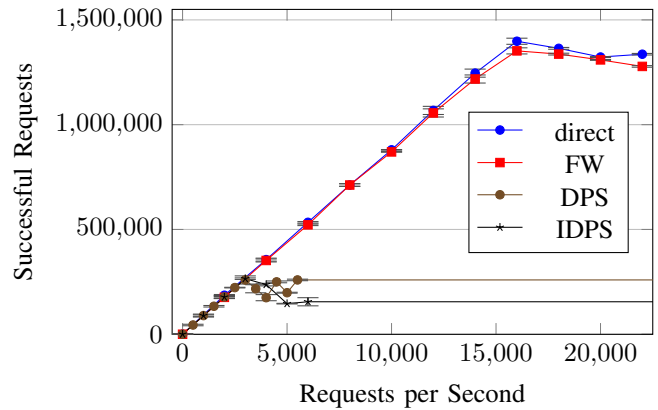

\begin{table}[!tb]
	\centering
	\begin{subtable}{\columnwidth}
	    \begin{center}
        	\begin{tabularx}{.98\columnwidth}{Xrrrr}
        		\toprule
        		Chain & \multicolumn{2}{c}{Average ICMP} & \multicolumn{2}{c}{Average SYN}\\ 
        		& response & packet loss & response & packet loss\\ \midrule
        		\midrule
        		Direct & 0.191ms & 0\% & 4.5ms & 0\% \\
        		Firewall & 0.343ms & 0\% & 3.8ms & 0\% \\
        		DPS & 0.194ms & 0\% & 4.7ms & 0\% \\
        		IDPS & 0.340ms & 0\% & 4.4ms & 0\% \\
        		\bottomrule
            \end{tabularx}
        	\caption{Single security functions with a benign workload.}
        	\label{tab:dynamic-reordering:ordering-effect:single-application-performance:benign:latency}
    	\end{center}
	\end{subtable}
	\begin{subtable}{\columnwidth}
		\vspace{5pt}
		\begin{center}
		\begin{tabularx}{.98\columnwidth}{ Xrrcc }
			\toprule
			Chain & \multicolumn{2}{c}{Average ICMP} & \multicolumn{2}{c}{\textcolor{white}{Average SYN}}\\ 
			& response & packet loss & \textcolor{white}{response} & \textcolor{white}{packet loss}\\ \midrule
			Firewall\( \rightarrow \)\ac{IDPS} & 0.390ms & 0\% & & \\ 
			\ac{IDPS}\( \rightarrow \)Firewall & 0.403ms & 0\% & &\\
			\bottomrule
		\end{tabularx}
		\caption{HTTP flood with firewall and \ac{IDPS}.}
		\label{tab:dynamic-reordering:ordering-effect:ssfc-performance:http-flood:firewall-vs-idps:stats}
	    \end{center}
	\end{subtable}
	\begin{subtable}{\columnwidth}
		\vspace{5pt}
		\begin{center}
		\begin{tabularx}{.98\columnwidth}{ Xrrrr }
			\toprule
			Chain & \multicolumn{2}{c}{Average ICMP} & \multicolumn{2}{c}{Average SYN}\\ 
			& response & packet loss & response & packet loss\\ \midrule
			\ac{DPS}\( \rightarrow \)Firewall & 0.288ms & 0\% & 1.0ms & 0\% \\
			Firewall\( \rightarrow \)\ac{DPS} & 0.242ms & 0\% & 4.5ms & 0\% \\
			\bottomrule
		\end{tabularx}
		\caption{SYN flood with \ac{DPS} and firewall.}
		\label{tab:dynamic-reordering:ordering-effect:ssfc-performance:syn-flood:dps-vs-firewall:stats}
	\end{center}
	\end{subtable}
	\begin{subtable}{\columnwidth}
    	\vspace{5pt}
		\begin{center}
		\begin{tabularx}{.98\columnwidth}{ Xrrrr }
			\toprule
			Chain & \multicolumn{2}{c}{Average ICMP} & \multicolumn{2}{c}{Average SYN}\\ 
			& response & packet loss & response & packet loss\\ \midrule
			\ac{IDPS}\( \rightarrow \)Firewall & 36.0ms & 30\% & 37.0ms & 34\% \\
			Firewall\( \rightarrow \)\ac{IDPS} & 16.0ms & 0\% & 17.9ms & 0\% \\
			\bottomrule
		\end{tabularx}
		\caption{Intrusion flood with \ac{IDPS} and firewall.}
		\label{tab:dynamic-reordering:ordering-effect:ssfc-performance:intrusion-flood:firewall-vs-idps:stats}
    	\end{center}
	\end{subtable}
	\caption{Latencies and packet losses for various configurations.}
\end{table}

Before evaluating the impact of the \ac{SSFC} order on performance, it is necessary to establish a baseline that establishes a realistic maximum performance that is attainable by the service host.
\Cref{fig:benign:overview} visualizes that the service scales linearly and beginning with 16000 requests/second the number of successful requests stalls. 
There is even a small decrease in throughput afterward which is probably attributable to queuing, swapping, and context switching effects. 
At that point, for the Direct Chain (both servers directly connected to the same switch), the target service has reached its limit.
The first data row in \Cref{tab:dynamic-reordering:ordering-effect:single-application-performance:benign:latency} shows very low latency and no packet loss. 
These results serve as a baseline for evaluating how the appliances impact the latency and data loss.

We repeat the same experiment with each single security function present and \Cref{fig:benign:overview} shows that the security functions behave differently.
The firewall closely follows the results from the Direct Chain.
The \ac{DPS} and the \ac{IDPS} both can follow only the direct connection up to 3000 requests/second and then stagnate or even lose performance.

When asserting the further metrics from \Cref{tab:dynamic-reordering:ordering-effect:single-application-performance:benign:latency}, we realize, that the firewall and the \ac{IDPS} increase the \ac{ICMP} response.
These results and further experiments lead us to the assumption that this effect roots in the number of necessary switches the packets must travel through.
Here, the \ac{DPS} connects via a single extra switch, while the firewall and the \ac{IDPS} connect to switches further up in the chain, creating additional hops.
For the SYN response the sole outlier is the firewall, that reduces this time.
No configuration creates packet loss for either \ac{ICMP} or SYN packets.

\subsection{Security Service Function Chain Performance}
\label{sec:dynamic-reordering:ordering-effect:ssfc-performance}

After evaluating the single security functions, we now proceed to combinations of security functions.
We create simulated attacks and combine pairs of security functions for each attack and switch their ordering for comparison.

\subsubsection{HTTP Flood}

The first benchmark is an \ac{HTTP} flood attack.
This attack aims to exhaust a services resources by issuing \ac{HTTP} requests either in high frequency or by targeted requests creating a high compute load.
The firewall is the defending security function and blocks \ac{HTTP} requests from the malicious sources.
We scale the \ac{HTTP} flood attack in steps of 1000 requests/second up to 14000 requests/second and perform 2000 benign \ac{HTTP} requests/second for one minute to assert the performance for benign workloads.
We measure metrics -- other than the throughput -- at 5000 requests/second flood strength.
We compare the following two \ac{SSFC} orders: 
\begin{enumerate*}[label=\emph{\alph*})]
    \item \ac{IDPS}\( \rightarrow \)Firewall, and
        \item Firewall \( \rightarrow \) \ac{IDPS}
\end{enumerate*}.

\begin{figure}[!tb]
	\centering
	\begin{tikzpicture}
    \begin{axis}[
        width=.9\columnwidth,
        height=.25\textheight,
        scaled ticks=false, 
        tick label style={/pgf/number format/fixed},
        ymajorgrids = true,
        ylabel= {Successful Requests},
        ymin = 0, ymax = 135000,
        ytick distance = 20000,
        xlabel = {HTTP Flood strength in Requests/s},
        xmin = 0, xmax = 14500,
        xtick distance = 2500,
        legend style=
        {
            column sep = 10pt,
            nodes=
            {
                inner sep=2, 
                below=-1.1ex
            }, 
            at={(0.95,0.6)}, 
            anchor=east
        }, 
    ]
        \addplot+[
            mark options={scale=.8},
            error bars/.cd, y dir=both, y explicit,
            error mark options = {
                rotate = 90, 
                mark size = 4pt, 
            }
        ]
        table[x=Load, y=IDPS_FW, y error = IDPS_FW_Error, col sep=comma]{data/05_dynamic-reordering/ordering-effect/ssfc-performance_http-flood_firewall-vs-idps_throughput.csv};
	\addlegendentry{IDPS\( \,\rightarrow\, \)Firewall}
        \addplot+[
            mark options={scale=.8},
            error bars/.cd, y dir=both, y explicit,
            error mark options = {
                rotate = 90, 
                mark size = 4pt, 
            }
        ]
        table[x=Load, y=FW_IDPS, y error = FW_IDPS_Error, col sep=comma]{data/05_dynamic-reordering/ordering-effect/ssfc-performance_http-flood_firewall-vs-idps_throughput.csv};
	\addlegendentry{Firewall\( \,\rightarrow\, \)IDPS}
    \end{axis}
\end{tikzpicture}
	\caption{Successful requests during an \ac{HTTP} flood attack.}
	\label{fig:dynamic-reordering:ordering-effect:ssfc-performance:http-flood:firewall-vs-idps:throughput}
\end{figure}
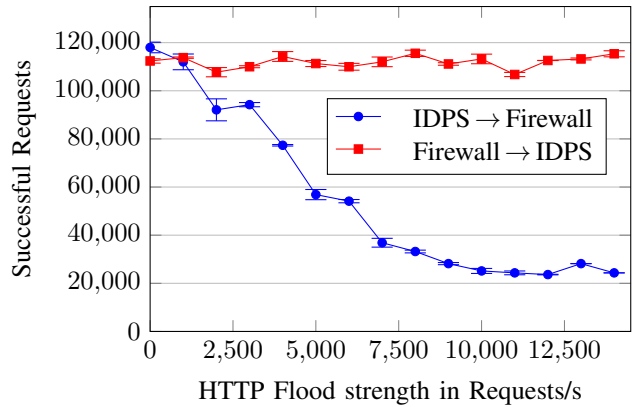

\Cref{fig:dynamic-reordering:ordering-effect:ssfc-performance:http-flood:firewall-vs-idps:throughput} presents the throughput results.
Both systems handle the benign workload and a small attack load of 1000 requests/second well. 
At higher attack load levels, the number of successful benign requests drops for the \ac{IDPS}\( \rightarrow \)Firewall \ac{SSFC} and from 10000 requests onwards settles just above 20000 successful requests.
Meanwhile, the Firewall\( \rightarrow \)\ac{IDPS} \ac{SSFC} is hardly affected by the attack load and remains close to the maximum attainable level and always stays significantly above the other \ac{SSFC} order's level.


\Cref{tab:dynamic-reordering:ordering-effect:ssfc-performance:http-flood:firewall-vs-idps:stats} shows further metrics for the two \ac{SSFC} orders. 
Considering these values, both \ac{SSFC} orders perform similarly.
When considering the \ac{CPU} load measurements (not shown), we see that the \ac{IDPS} is at maximum load for both \ac{SSFC} orders. 
However, when the firewall is not in front, at the beginning, a clear overload is visible.
The firewall resides at very low load levels and shows to have significant reserves.

\subsubsection{\acs{SYN} Flood}
\label{sec:dynamic-reordering:ordering-effect:ssfc-performance:syn-flood}

As a second benchmark, we perform a SYN Flood attack.
This attack aims at exhausting a server's buffer for half-open \ac{TCP} connections. 
The \ac{DPS} is the defending security function.
For each run, we increase the SYN flood strength by 500~Mbit/s, up to 6500~Mbit/s. 
We generate a load of 2000 benign HTTP requests/second for one minute to evaluate the successful requests during a SYN flood.
We measure metrics other than the successful requests at 5000~MBit/s attack load.
For this attack, we benchmark two more \ac{SSFC} orders: 
\begin{enumerate*}[label=\emph{\alph*})]
    \item \ac{DPS}\( \rightarrow \)Firewall, and
    \item Firewall \( \rightarrow \) \ac{DPS}
\end{enumerate*}.

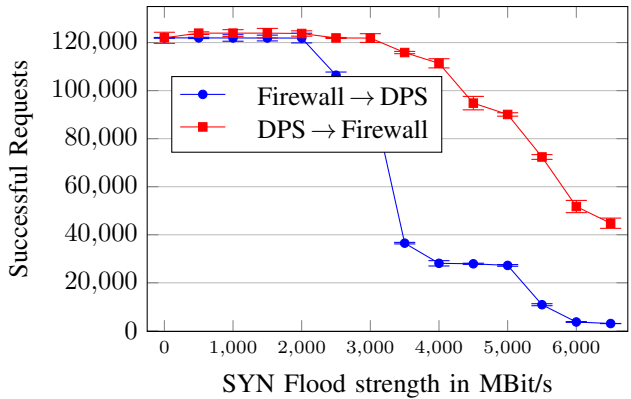
\begin{figure}[!tb]
	\centering
	\begin{tikzpicture}
    \begin{axis}[
        width=.9\columnwidth,
        height=.25\textheight,
        scaled ticks=false, 
        tick label style={/pgf/number format/fixed},
        ymajorgrids = true,
        ylabel= {Successful Requests},
        ymin = -250, ymax = 135000,
        ytick distance = 20000,
        xlabel = {SYN Flood strength in MBit/s},
        xmin = -250, xmax = 6750,
        x tick label style = {font=\scriptsize},
        legend style=
        {
            column sep = 10pt,
            nodes=
            {
                inner sep=2, 
                below=-1.1ex
            }, 
            at={(0.05,0.55)}, 
            anchor=south west
        }, 
    ]
        \addplot+[
            mark options={scale=.8},
            error bars/.cd, y dir=both, y explicit,
            error mark options = {
                rotate = 90, 
                mark size = 4pt, 
            }
        ]
	table[x=Load, y=FW_DPS, y error = FW_DPS_Error, col sep=comma]{data/05_dynamic-reordering/ordering-effect/ssfc-performance_syn-flood_dps-vs-firewall_throughput.csv};
	\addlegendentry{Firewall\( \,\rightarrow\, \)DPS}
        \addplot+[
            mark options={scale=.8},
            error bars/.cd, y dir=both, y explicit,
            error mark options = {
                rotate = 90, 
                mark size = 4pt, 
            }
        ]
	table[x=Load, y=DPS_FW, y error = DPS_FW_Error, col sep=comma]{data/05_dynamic-reordering/ordering-effect/ssfc-performance_syn-flood_dps-vs-firewall_throughput.csv};
	\addlegendentry{DPS\( \,\rightarrow\, \)Firewall}
    \end{axis}
\end{tikzpicture}
	\caption{Successful requests during a SYN flood attack.}
	\label{fig:dynamic-reordering:ordering-effect:ssfc-performance:syn-flood:dps-vs-firewall:throughput}
\end{figure}
	
\Cref{fig:dynamic-reordering:ordering-effect:ssfc-performance:syn-flood:dps-vs-firewall:throughput} presents the number of successful requests for both \ac{SSFC} orders, which keep an optimal success rate until 2000~MBit/s.
After that, the firewall-headed \ac{SSFC} order slowly drops. There is a very steep drop at 3500~MBit/s.
At 6000~MBit/s, it further drops to 3688, and then stays at a similar level.
The \ac{DPS}\( \rightarrow \)Firewall \ac{SSFC} continues at maximum performance until 3000~MBit/s.
Then, it continuously drops but stays significantly above the inverse chain's performance.

		
\Cref{tab:dynamic-reordering:ordering-effect:ssfc-performance:syn-flood:dps-vs-firewall:stats} shows further metrics for both \ac{SSFC} orders.
The results are a little more diverse than for the previous attack and combinations.
The Firewall\( \rightarrow \)\ac{DPS} \ac{SSFC} offers a faster \ac{ICMP} response while the \ac{DPS}\( \rightarrow \)Firewall \ac{SSFC} yields a faster SYN response.
The relative difference for the SYN response is larger than for the \ac{ICMP} response.
Both configurations do not yield packet losses.

The Firewall\( \rightarrow \)\ac{DPS} \ac{SSFC} impacts the load during the attack.
The background load (e.g., \acs{OS} operations or filesystem journaling) in the \texttt{user} and \texttt{system} state is forced out by the actual load of the firewall application.
This application load appears in the \texttt{softirq} level, where the \ac{CPU} spends 100\% of its time.
When reversing the \ac{SSFC} order, no noticeable load shows in the \texttt{softirq} state, and the background loads remain in the \texttt{user} and \texttt{system} state.
Thus, this \ac{SSFC} order eliminates all load on the firewall.

\subsubsection{Intrusion Flood}
\label{sec:dynamic-reordering:ordering-effect:ssfc-performance:intrusion-flood}

The intrusion flood attack aims at abusing vulnerabilities inside a service.
We use \ac{UDP} packets containing a signature that matches the \ac{IDPS} rules to create the flood and perform the intrusion flood for up to 5000~MBit/s scaling in steps of 500~MBit/s.
Further, we measure further metrics at an attack load of 1000~MBit/s and compare the same \ac{SSFC} orders as for the \ac{HTTP} flood.

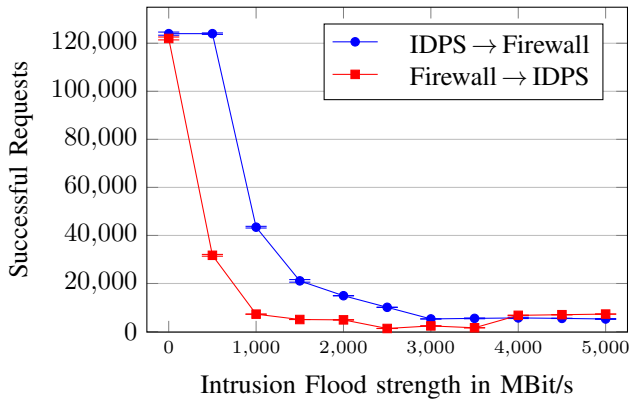
\begin{figure}[!tb]
	\centering
	\begin{tikzpicture}
    \begin{axis}[
        width=.9\columnwidth,
        height=.25\textheight,
        scaled ticks=false, 
        tick label style={/pgf/number format/fixed},
        ymajorgrids = true,
        ylabel= {Successful Requests},
        ymin = -250, ymax = 135000,
        ytick distance = 20000,
        xlabel = {Intrusion Flood strength in MBit/s},
        xmin = -250, xmax = 5250,
        x tick label style = {font=\scriptsize},
        legend style=
        {
            column sep = 10pt,
            nodes=
            {
                inner sep=2, 
                below=-1.1ex
            }, 
            at={(0.95,0.95)}, 
            anchor=north east
        }, 
    ]
        \addplot+[
            mark options={scale=.8},
            error bars/.cd, y dir=both, y explicit,
            error mark options = {
                rotate = 90, 
                mark size = 4pt, 
            }
        ]
        table[x=Load, y=IDPS_FW, y error = IDPS_FW_Error, col sep=comma]{data/05_dynamic-reordering/ordering-effect/ssfc-performance_intrusion-flood_firewall-vs-idps_throughput.csv};
	\addlegendentry{IDPS\( \,\rightarrow\, \)Firewall}
        \addplot+[
            mark options={scale=.8},
            error bars/.cd, y dir=both, y explicit,
            error mark options = {
                rotate = 90, 
                mark size = 4pt, 
            }
        ]
        table[x=Load, y=FW_IDPS, y error = FW_IDPS_Error, col sep=comma]{data/05_dynamic-reordering/ordering-effect/ssfc-performance_intrusion-flood_firewall-vs-idps_throughput.csv};
	\addlegendentry{Firewall\( \,\rightarrow\, \)IDPS}
    \end{axis}
\end{tikzpicture}
	\caption{Successful requests during an intrusion flood attack.}
	\label{fig:dynamic-reordering:ordering-effect:ssfc-performance:intrusion-flood:firewall-vs-idps:throughput}
\end{figure}

\Cref{fig:dynamic-reordering:ordering-effect:ssfc-performance:intrusion-flood:firewall-vs-idps:throughput} shows the number of successful requests for both \ac{SSFC} orders.
Already at a flood strength of 500~MBit/s the Firewall\( \rightarrow \)\ac{IDPS} chain drops to 31732 successful requests. 
At 1000~MBit/s flood strength this chain further drops to 7238 successful requests and from thereon stays at similar or lower levels.
The reverse chain's performance drops later, starting at 1000~MBit/s with a drop to 43442~MBit/s.
It then continues to slowly fall and finally aligns with the firewall-headed chains throughput at 4000~MBit/s.
Between the beginning of attacks at 500~MBit/s and the alignment, the \ac{IDPS}\( \rightarrow \)Firewall \ac{SSFC} outperforms the other chain.


The \ac{IDPS}\( \rightarrow \)Firewall \ac{SSFC} about doubles the response time compared to its counterpart (see \Cref{tab:dynamic-reordering:ordering-effect:ssfc-performance:intrusion-flood:firewall-vs-idps:stats}.
It also introduces a packet loss rate of about one third of packets. 
This result surprises, since the higher throughput of the \ac{IDPS}-headed chain did not hint at this behavior.
However, a way of getting higher throughput might lie in accepting packet losses.
Putting the firewall first creates \texttt{user} and \texttt{system} load for both systems. 
While the firewall is not in an extremely high load situation, the \ac{IDPS} is in overload.
Changing the \ac{SSFC} order results in taking away the load from the firewall and the \texttt{system} load from the \ac{IDPS} but heavily overloads the \ac{IDPS}.
The firewall spends most of its time the \texttt{softirq} state, when it is first in the chain.
However, when the \ac{IDPS} heads the chain, only a small peak appears at the beginning.

\subsection{Discussion}
\label{sec:dynamic-reordering:ordering-effect:discussion}

\Cref{sec:dynamic-reordering:ordering-effect:single-application-performance} shows that even under benign workloads, the different security functions perform with significant differences.
While the firewall can protect a service without reducing the throughput, the \ac{DPS} and the \ac{IDPS} reach their limits far before the protected service.
Also, both systems (the \ac{IDPS} more than the \ac{DPS}) show that their performance can further drop when the load increases further. 

\Cref{sec:dynamic-reordering:ordering-effect:ssfc-performance} confirms our assumption that the \ac{SSFC} order has a significant impact on \ac{SSFC} performance. 
When considering the throughput, we see different behaviors when comparing different attacks.
Those behaviors share one commonality -- placing the security function that defends against the attack first yields the most successful benign requests.
In some cases, the right \ac{SSFC} order significantly prolongs the load level at which performance drops and slows the drop. 
Still, at some point, both \ac{SSFC} orders converge to similar results.

We show that the \ac{SSFC} order has a significant effect on the throughput, other metrics and the \ac{CPU} load. 
For the selected attack combinations, we also find that there is no optimal \ac{SSFC} order for all attacks. While for \ac{HTTP} flood, the firewall performs best before the \ac{IDPS}, the reversed chain is superior during an intrusion flood.
In general, putting the security function dedicated to protecting against the current attack first, yields the best results.
Therefore, we require different \ac{SSFC} orders depending on the current attack state of the system.
This finding confirms our claim that dynamic \ac{SSFC} reordering can improve the performance of \acp{SSFC}. 
We will follow the realization of this concept in the following sections.

\section{A Framework for Attack-aware Security Function Chain Reordering}
\label{sec:framework}

In this section, we will present an architecture for an attack-aware dynamic \ac{SSFC} reordering framework and provide a \ac{PoC} implementation.
We then evaluate this implementation and its capabilities and discuss the results and further challenges.

\subsection{Architecture}
\label{sec:dynamic-reordering:framework:architecture}

\begin{figure}[!tb]
    \centering
    \includegraphics[width=.96\columnwidth]{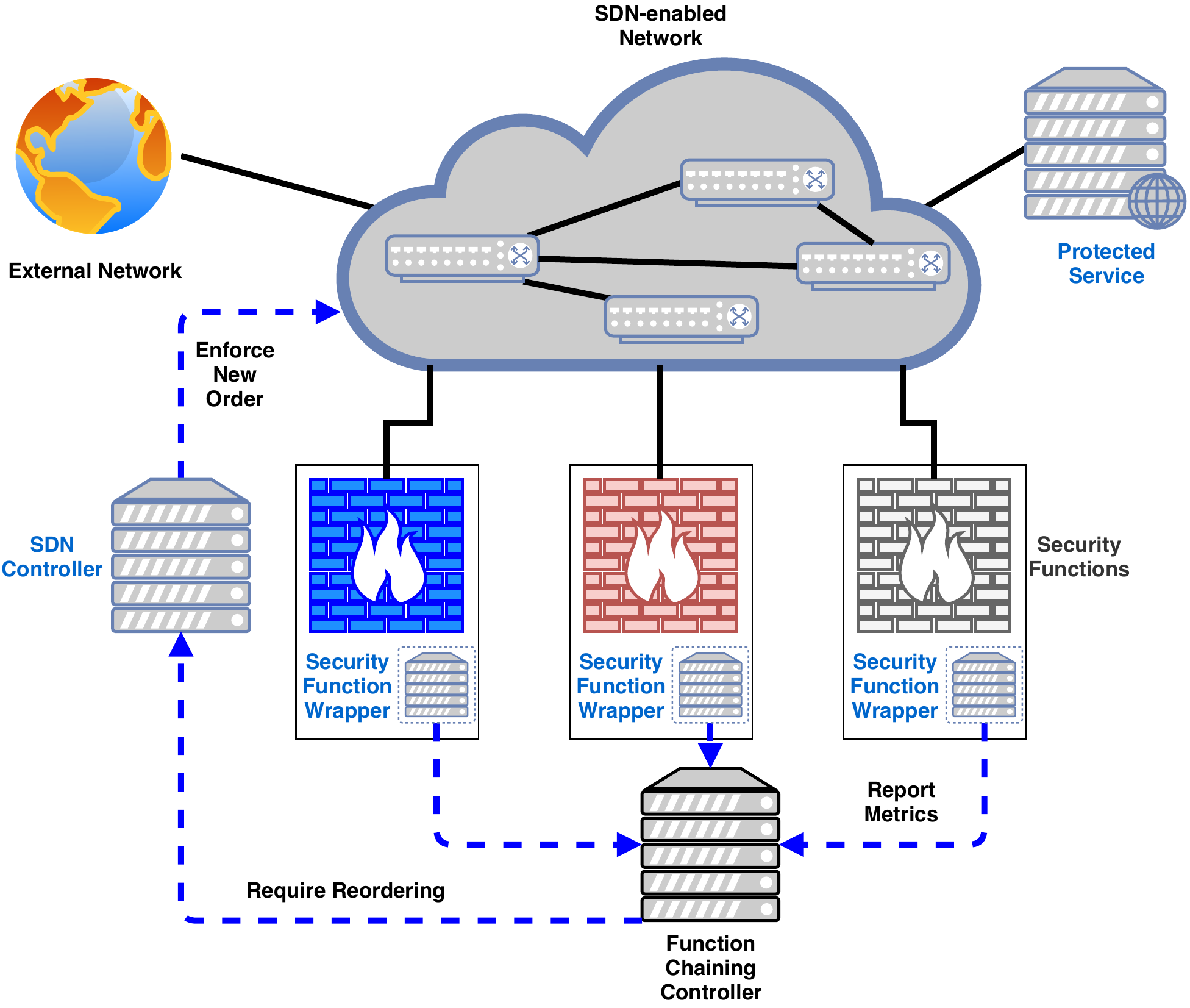}
    \caption{Components of the attack-aware \ac{SSFC} reordering framework.}
    \label{fig:dynamic-reordering:framework:architecture:components}
\end{figure}

The attack-aware \ac{SSFC} reordering framework consists of multiple components, as depicted in \Cref{fig:dynamic-reordering:framework:architecture:components}.
A generic \ac{SDN}-enabled network connects the external network and the service protected by our security system.
All relevant security functions connect to that network as well.
We deploy so-called security function wrappers alongside the security functions to gather metrics about them and their attack and report them to the \ac{FCC}.
The \ac{FCC} collects data from the wrappers and optionally other sources. 
It forms a decision whether an \ac{SSFC} reordering is necessary and in that case sends the new \ac{SSFC} order to the \ac{SDN} controller which enforces the new order inside the \ac{SDN}-enabled network.

The \emph{Security Function Wrapper}
is a program running on the security function hosts and communicates with the \ac{FCC}. 
It is responsible for registering the security function at the \ac{FCC}, deleting it on graceful shutdowns, keeping a connection to the \ac{FCC} to allow the management of security functions through the \ac{FCC}, and finally, offering an interface for the security function to report detected attacks to the \ac{FCC} over the wrapper.
At first, the security function wrapper validates and loads its configuration.
If everything is loaded correctly, it registers with the \ac{FCC} and receives a token for further communication.
In a keep-alive loop, the security function wrapper periodically sends a keep-alive message to the \ac{FCC} and receives an updated token.
In its main loop, it reports attacks registered by the monitored security function after performing a validation (e.g., a machine on a 1GBit/s interface could not report a valid attack with a strength of 5~GBit/s) to the \ac{FCC}.
Upon shutdown, it deregisters with the \ac{FCC}.

The \emph{\acl{FCC}} runs in a centralized location reachable from all security functions.
A webpage showing attack statistics and the current and standard configuration is part of the \ac{FCC}. 
Additionally, the webpage contains a form to change the routing configuration manually based on the available groups of security functions. 
The controller needs to handle the requests from the wrapper instance, namely registration, delete requests, attack alerts, and keep-alive requests, as shown before.
The \ac{FCC} must keep a list of the security function groups and their respective attack rate to calculate the new optimal routing configuration reactively. 
After calculating the new routing configuration, the \ac{FCC} sends it to the \ac{SDN} controller, which then applies it to the switches and, therefore, the network.
In this section, we use a simplified approach putting the security function group with the most attacks at the front.
After successfully changing the routing configuration, the \ac{FCC} changes the stored current configuration to the new routing configuration and resets the reported attacks.

\subsection{Evaluation}
\label{sec:dynamic-reordering:framework:evaluation}

\subsubsection{Implementation and Evaluation Environment}

The complete framework uses Python~3 with four non-standard Python libraries: Flask, PyJWT, requests, and netifaces.
In general, this framework supports every \ac{SDN} controller offering a \ac{REST} \ac{API} for flow modification.
To ensure the absence of side effects from the \ac{SDN} controller, we implemented a minimalistic \ac{SDN} controller ourselves.
For this \ac{PoC} evaluation and the following evaluation, we limit our framework to use Open vSwitch only.
This \ac{SDN} Controller consists of two \emph{Flask} applications: the actual controller and a switch wrapper running on the \textbf{Open vSwitch} machines. 

We use a testbed environment similar to the one presented before for the \ac{SSFC} order evaluation.
However, we replace the physical switches with Open vSwitch instances.

\subsubsection{Manual Reordering}
\label{sec:dynamic-reordering:framework:evaluation:manual-reconfiguration}

At first, we evaluate that our system is able to correctly apply reordering decisions.

\paragraph{Experiment Description}

We test all six \ac{SSFC} orders possible for the three security functions.
Therefore, we automated the process to test these routing configurations in the network based on the standard configuration. 
After starting the system and registering the security functions, the \emph{client} sends an \ac{ICMP} echo-request to the \emph{server}.
Next, we started \emph{tcpdump} on each security function logging the traffic. 
From these logs, we construct the path, a packet takes through the network.

\paragraph{Results}

The results show that our framework applies every permutation of the default configuration correctly. 
When reordering, it is possible to traverse through more security functions than there are in the network. 
This issue is a result of changing the routing configuration while traffic passes through the system. 
The following packets go through the desired function chain. 
Although packet loss is theoretically possible, the new routing configurations are applied instantly for the retransmissions. 
Also, while theoretically possible, no attack completely skipped a security function. 

\subsubsection{Reaction to Simulated Attacks}
\label{sec:dynamic-reordering:framework:evaluation:simulated-attacks}

Next, we analyze how the system reacts to simulated attacks.

\paragraph{Experiment Description}

This experiment validates whether the \ac{FCC} correctly changes the routing configuration based on the attacks reported by the security functions.
The main idea that led to the development of the framework is to change the routing configuration \emph{dynamically}. 
As described before, the security functions report detected attacks via the co-located security function wrapper instance to the \ac{FCC}. 
We simulate attacks on each \ac{VM}, to show that the attack reporting works and the routing configuration changes depending on the attack reports.
The simulated security functions send attack reports with a changing probabilities.
We configured a threshold of \(100\) attacks in the \ac{FCC}. 
Only if the attack count exceeds this threshold, the \ac{FCC} calculates and --- if necessary --- applies a new routing configuration. 
Additionally, we define an \emph{imminent threshold} three times larger than the regular threshold checked every ten seconds.

\paragraph{Experiment Results}

\begin{figure}[!tb]
    \centering
    \includegraphics[width=.96\columnwidth]{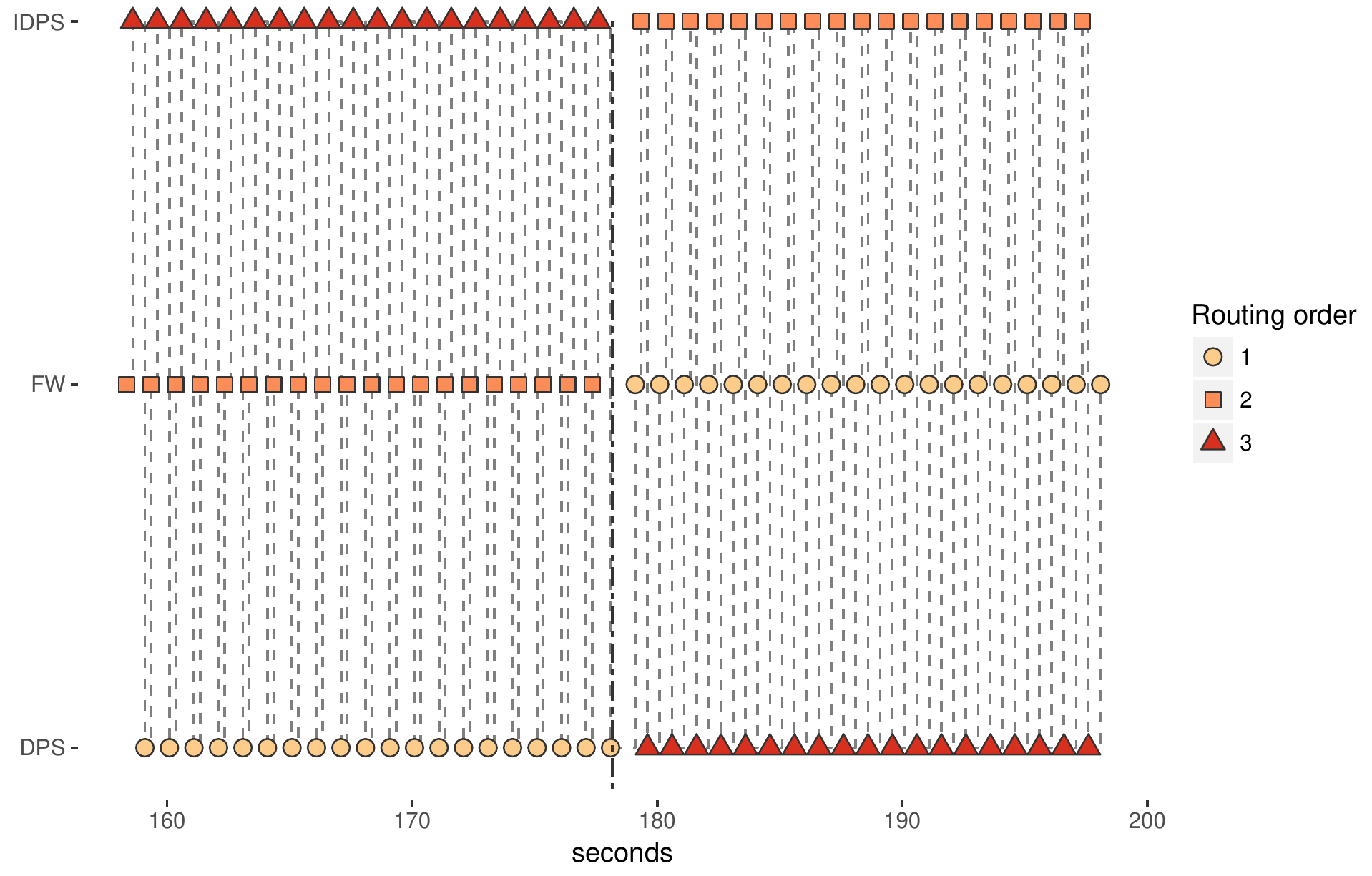}
    \caption{Change of routing configuration from \ac{DPS}-Firewall-\ac{IDPS} to Firewall-\ac{IDPS}-\ac{DPS}}
    \label{fig:dynamic-reordering:framework:evaluation:simulated-attacks:change1}
\end{figure}

\Cref{fig:dynamic-reordering:framework:evaluation:simulated-attacks:change1} illustrates the \emph{imminent attack} functionality. 
The first modification of the routing configuration occurs at almost three minutes into the experiment. 
The \ac{SSFC} order changes from \emph{\ac{DPS}-Firewall-\ac{IDPS}} to \emph{Firewall-\ac{IDPS}-\ac{DPS}}. 
This change could not originate from the regular check as it occurs before the five-minute mark. 

A further change proves the regular functionality of the \ac{FCC}. 
The routing configuration takes effect approximately 12 minutes after the start of the experiment. 
The \ac{SSFC} order changes from \emph{Firewall-\ac{IDPS}-\ac{DPS}} to \emph{\ac{IDPS}-Firewall-\ac{DPS}}. 
Here, it becomes visible that for one ping, the packets did not traverse every security function, posing a potential security risk. 

Further reorderings work as desired.
The \ac{FCC} resets the configuration to the standard configuration if reported attacks of all security functions are below the configured threshold.
This reset to default allows users to select a default configuration that best fits the average attack on the system.

\subsubsection{Discussion}
\label{sec:dynamic-reordering:framework:discussion}

\paragraph{Functionality}

In summary, the developed framework is working as expected.
Small issues like packet loss may occur during the application of new routing configurations. 
The generation of routes and their application works as desired.
We also showed that the framework is indeed attack-aware and successfully changes the routing configuration of the network based on the reported attacks from the security functions. 
After attacks fade out, the framework then switches back to the default configuration.

\paragraph{Security Issues During Reconfiguration}

Three undesired scenarios can occur during reconfiguration:
\begin{enumerate*}[label=\emph{\roman*})]
    \item packets get dropped because the framework has not installed the required flow yet,
    \item packets traverse through more than once through one or more security functions, and
    \item packets do not traverse through all required security functions.
\end{enumerate*}

The first and second issues pose only \acl{QoS} and \acl{QoE} concerns. 
However, the third issue is relevant to security. 
If packets can skip security functions, single malicious packets can reach the receiver.
This issue is of little concern for flood attacks, but for intrusions, a single packet might be enough to trigger a vulnerability and cause a severe security breach. 

To avoid this issue, we propose several solutions:
\begin{enumerate*}[label=\emph{\roman*})]
    \item A second set of security functions.
        Reconfigurations would then use this second set for the \ac{SSFC}.
        Once all packets clear the security functions in the first chain, those functions become the spare functions for the next reconfiguration.
    \item To model the stay of packets inside the security function, by adding short-lived flows with artificial delays that ensure, that no packets are inside the functions when executing the reordering.
        However, this requires detailed knowledge of all security functions inside the chain, especially regarding their queuing behavior. 
    \item To force the security functions to drop all packets before executing the reordering.
        This solution fixes the second and third problems but moves the affected packets and others to the first issue.
    \item To use the options field in the \ac{IP} header. 
        We create a counter field in the options and increment it for every reconfiguration.
        The inbound switch has a rule that modifies incoming packet headers to contain the current counter value, and all created flows match against the current counter.
        Older flows expire after some time.
\end{enumerate*}

Depending on the use-case, there are different optimal solutions for the security system architecture.
We have not yet implemented these solutions but will do so in the future. 
For the \ac{PoC}, the presented implementation is sufficient.
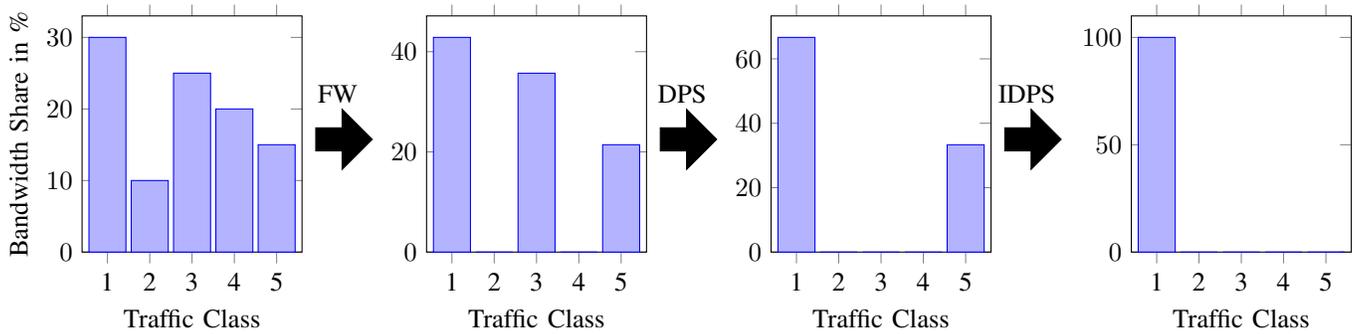
\begin{figure*}[!tb]
    \centering
    \begin{tikzpicture}
    \begin{axis}[
        ybar,
        height = .2\textheight,
        bar width = 14pt,
        x = 16pt,
        ymin = 0,
        enlarge x limits=0.15,
        ylabel={Bandwidth Share in \%},
        symbolic x coords={1,2,3,4,5},
        xtick=data,
        xlabel={Traffic Class},
        nodes near coords align={vertical},
        ]
    \addplot coordinates {(1,30) (2,10) (3,25) (4,20) (5,15)};
    \end{axis}
    \node[draw, single arrow,
          minimum height=7.5mm, minimum width=8mm,
          single arrow head extend=2mm,
          anchor=west, rotate=0, fill = black] at (3.1,1.5) {};
    \node at (3.4,2.1) {FW};
\end{tikzpicture}
\begin{tikzpicture}
    \begin{axis}[
        ybar,
        height = .2\textheight,
        bar width = 14pt,
        x = 16pt,
        ymin = 0,
        enlarge x limits=0.15,
        symbolic x coords={1,2,3,4,5},
        xtick=data,
        xlabel={Traffic Class},
        nodes near coords align={vertical},
        ]
    \addplot coordinates {(1,42.86) (2,0) (3,35.71) (4,0) (5,21.43)};
    \end{axis}
    \node[draw, single arrow,
        minimum height=7.5mm, minimum width=8mm,
        single arrow head extend=2mm,
        anchor=west, rotate=0, fill = black] at (3.1,1.5) {};
    \node at (3.4,2.1) {DPS};
\end{tikzpicture}
\begin{tikzpicture}
    \begin{axis}[
        ybar,
        height = .2\textheight,
        bar width = 14pt,
        x = 16pt,
        ymin = 0,
        enlarge x limits=0.15,
        symbolic x coords={1,2,3,4,5},
        xtick=data,
        xlabel={Traffic Class},
        nodes near coords align={vertical},
        ]
        \addplot coordinates {(1,66.67) (2,0) (3,0) (4,0) (5,33.33)};
    \end{axis}
    \node[draw, single arrow,
        minimum height=7.5mm, minimum width=8mm,
        single arrow head extend=2mm,
        anchor=west, rotate=0, fill = black] at (3.1,1.5) {};
    \node at (3.4,2.1) {IDPS};
\end{tikzpicture}
\begin{tikzpicture}
    \begin{axis}[
        ybar,
        height = .2\textheight,
        bar width = 14pt,
        x = 16pt,
        ymin = 0,
        enlarge x limits=0.15,
        symbolic x coords={1,2,3,4,5},
        xtick=data,
        xlabel={Traffic Class},
        nodes near coords align={vertical},
        ]
    \addplot coordinates {(1,100) (2,0) (3,0) (4,0) (5,0)};
    \end{axis}
\end{tikzpicture}
    \caption{Exemplary development of the traffic composition over the course of a security service function chain.}
    \label{fig:dynamic-reordering:modeling:function-chain:traffic:example}
\end{figure*}

\section{Modeling Security Function Chains}
\label{sec:dynamic-reordering:modeling}

Here, we will develop a precise model.
First, we take a look at how to model a single security function based on the incoming traffic.
We then use this knowledge and combine multiple security function models into a single \ac{SSFC} model.

\subsection{Modeling Single Security Functions}
\label{sec:dynamic-reordering:modeling:single-functions}

\subsubsection{Modeling Traffic}

Different security functions show a different behavior under different types of traffic.
Thus, we need a model that takes both benign traffic and the various attack types into account.
To this end, our model of the arrival rate must consider the content and the composition of the traffic.
Therefore, we model the traffic as different workload classes.
For every workload class, we record the rate of packets and the used bandwidth.
We model the traffic composition for the link from an external network to the first security function, every connection between security functions, and the link from the last security function to the protected system.

\subsubsection{Security Function Modeling}

With a model for the traffic, it is possible to model the behavior of the single security functions.
We propose to apply architectural performance models to model security functions.
Architectural models capture the semantics, allowing for a plain view on the security functions, in contrast to low-level stochastic formalisms.
We model each security function as a software component.
However, we also offer simplified approaches to model security functions.
It is necessary to model three aspects: 
\begin{enumerate*}[label=\emph{\roman*})]
    \item the effect of the security functions on the traffic composition, 
    \item the performance behavior of the security function, and
    \item tertiary effects like packet-loss
\end{enumerate*}.

Based on the distribution of the input traffic of a security appliance, the corresponding output traffic can be derived. 
We define the distribution of the input/output traffic as \(P_{in/out}(t_i)\) with \(i \in [1,n]\) for n different types of traffic. 

Exemplary, for a security function which drops all packets of the traffic type \(k\) the output traffic looks as follows: 
\begin{equation}
P_{out}(t_i) = \begin{cases} 
    P_{in}(t_i)/(1 - P_{in}(t_k))     &   \mbox{for } i \neq k \\ 
    0       &   \mbox{for } i = k 
\end{cases}.
\end{equation}
The current model assumes that a function eliminates all malicious traffic of one or more traffic types.


For \emph{performance modeling}, the most precise modeling solution would be to use a full-blown model of the software component to model the function's performance behavior.
Such models usually base on the functions source code or are extracted by heavy black-box testing.
However, often neither the source code nor the resources are available to perform extensive black-box testing.
Still, even when such a model is not possible, for our security functions, we have found two general types of resource demand generation. 
The first type is a constant demand created per unit (e.g., frame, packet, segment, request). 
The second type creates a demand correlating to the size of the unit with different correlation types. 


There are \emph{tertiary factors} that can increase the accuracy of the model in certain situations.
These factors do not correlate directly with single packets or the traffic distribution but instead with the state of the security.
Examples are 
    Queuing Behavior,
    False Positives and False Negatives,
    Drop Rate,
    Overload Behavior, and
    Short Term \ac{CPU} Frequency Scaling.

\subsection{Modeling Security Service Function Chains}
\label{sec:dynamic-reordering:modeling:function-chain}

Based on the model for single security functions, it is possible to model the whole security function chain.
Therefore, we model the chain by putting the functions one after the other and feeding the output of the previous function to the next.

When starting with the input traffic, the traffic results from putting it through one function after the other. 
\Cref{fig:dynamic-reordering:modeling:function-chain:traffic:example} shows such a development of the traffic for the function described above.
Therefore, we use a simplified model without tertiary factors.
The traffic starts with a distribution over all traffic types. 
At every security function, this function removes one or more traffic classes.
Thereby the share of the other classes increases.
This process repeats itself at every security function until only the benign traffic remains.
This combination allows a full model of the chain when knowing the composition at the startup.
However, most of the time, this composition is unknown.
Still, it is possible to reverse engineer this composition using the reports from the security function wrappers and the switches in the framework.

\section{Conclusion}
\label{sec:conclusion}

In this paper, we introduced the concept of attack-aware dynamic \acl{SSFC} reordering.
This concept incorporates changing the order of \acp{SSFC} to optimize them to most efficiently counter attacks.
At first, we described the general idea. 
The main component is the \acf{FCC}. 
It gathers information to model the security systems state, uses this information to compute the desired configuration, and enforces the \ac{SSFC} order. 
Next, we developed an evaluation environment for individual security functions and \acp{SSFC}.
When benchmarking single security functions, we found different types of behavior.
For every tested combination, the \ac{SSFC} order has a significant impact on the system's performance.
In general, putting the function that defends against the attack first, yields better performance.
The difference can make up two or more \ac{SSFC} orders of magnitude. 
For different attacks, we found \ac{SSFC} orders that contradicted each other.
Thus, there is no \ac{SSFC} order that is optimal for every attack.

Next, we have designed a framework that performs attack-aware dynamic \ac{SSFC} reordering.
All security functions reside inside an \ac{SDN}-enabled network.
A \emph{security function wrapper} co-located with every security function reports attacks at these functions via a separate management network to the \ac{FCC}.
The \ac{FCC} computes the desired \ac{SSFC} order for the security functions and submits it to the \ac{SDN} controller, which enforces it by creating the necessary flows on the \ac{SDN} switches.
We developed a \ac{PoC} implementation and show that the framework can enforce all possible \ac{SSFC} orders.
The framework successfully adapted to all attacks and, after the attacks ceased, successfully restored the default configuration.
Thus this proved the desired functionality.
An issue occurs when reordering, packets can drop or pass through a function twice. 
We proposed four options to combat this issue for different use-cases.

We model the traffic categorizing it into traffic classes, where benign traffic and every attack type each forms a class.
Every security function affects the traffic as a function depending on the traffic composition.
The model for an \ac{SSFC} consists of multiple security function models.
Traffic that exits one function continues to the next.
Thereby, it is possible to compute the total resource demand.

The work on dynamic \ac{SSFC} reordering is far from finished.
In the future, we plan to evaluate our modeling and test it with various decision-making approaches.
We also plan to extend the framework to allow different orders for different traffic types and evaluate its impact on energy consumption.

\bibliographystyle{IEEEtran}
\bibliography{dissertation,publications}

\begin{thebibliography}{10}
\providecommand{\url}[1]{#1}
\csname url@samestyle\endcsname
\providecommand{\newblock}{\relax}
\providecommand{\bibinfo}[2]{#2}
\providecommand{\BIBentrySTDinterwordspacing}{\spaceskip=0pt\relax}
\providecommand{\BIBentryALTinterwordstretchfactor}{4}
\providecommand{\BIBentryALTinterwordspacing}{\spaceskip=\fontdimen2\font plus
\BIBentryALTinterwordstretchfactor\fontdimen3\font minus
  \fontdimen4\font\relax}
\providecommand{\BIBforeignlanguage}[2]{{%
\expandafter\ifx\csname l@#1\endcsname\relax
\typeout{** WARNING: IEEEtran.bst: No hyphenation pattern has been}%
\typeout{** loaded for the language `#1'. Using the pattern for}%
\typeout{** the default language instead.}%
\else
\language=\csname l@#1\endcsname
\fi
#2}}
\providecommand{\BIBdecl}{\relax}
\BIBdecl

\bibitem{Vigna_2004_TestingNetwork-basedIntrusionDetectionSignaturesUsingMutantExploits}
G.~Vigna, W.~Robertson, and D.~Balzarotti, ``Testing network-based intrusion
  detection signatures using mutant exploits,'' in \emph{Proceedings of the
  11th ACM conference on Computer and communications security - CCS 04},
  ACM.\hskip 1em plus 0.5em minus 0.4em\relax ACM Press, 2004, pp. 21--30.

\bibitem{Oppliger_1997_CommunicationsoftheACM_InternetSecurityFirewallsandBeyond}
R.~Oppliger, ``Internet security: Firewalls and beyond,'' \emph{Communications
  of the {ACM}}, vol.~40, no.~5, pp. 92--102, May 1997.

\bibitem{Chiosi_2012_NetworkFunctionsVirtualizationNFVanIntroductionBenefitsEnablersChallenges&CallforAction}
\BIBentryALTinterwordspacing
M.~Chiosi, D.~Clarke, P.~Willis, A.~Reid, J.~Feger, M.~Bugenhagen, W.~Khan,
  M.~Fargano, D.~C. Cui, D.~H. Deng, J.~Benitez, U.~Micheel, H.~Damker,
  K.~Ogaki, T.~Matsuzaki, M.~Fukui, K.~Shimano, D.~Delisle, Q.~Loudier,
  C.~Kolias, I.~Guardini, E.~Demaria, R.~Minerva, A.~Manzalini, D.~Lopez,
  F.~J.~R. Salguero, F.~Ruhl, and P.~Sen, ``Network functions virtualization
  (nfv), an introduction, benefits, enablers, challenges \& call for action,''
  SDN and OpenFlow World Congress, Darmstadt, Germany, 2012. [Online].
  Available: \url{http://portal.etsi.org/NFV/NFV_White_Paper.pdf}
\BIBentrySTDinterwordspacing

\bibitem{Rygielski_2017_FlexibleModelingofDataCenterNetworksforCapacityManagement}
\BIBentryALTinterwordspacing
P.~Rygielski, ``Flexible modeling of data center networks for capacity
  management,'' Ph.D. dissertation, University of Würzburg, Germany, Mar.
  2017. [Online]. Available:
  \url{https://opus.bibliothek.uni-wuerzburg.de/frontdoor/index/index/docId/14623}
\BIBentrySTDinterwordspacing

\bibitem{Yan_2015_IEEECommunicationsMagazine_DistributedDenialofServiceAttacksinSoftware-definedNetworkingwithCloudComputing}
Q.~Yan and F.~R. Yu, ``Distributed denial of service attacks in
  software-defined networking with cloud computing,'' \emph{{IEEE}
  Communications Magazine}, vol.~53, no.~4, pp. 52--59, Apr. 2015.

\bibitem{Scott-Hayward_2016_IEEECommunicationsSurveys&Tutorials_ASurveyofSecurityinSoftwareDefinedNetworks}
S.~Scott-Hayward, S.~Natarajan, and S.~Sezer, ``A survey of security in
  software defined networks,'' \emph{{IEEE} Communications Surveys {\&}
  Tutorials}, vol.~18, no.~1, pp. 623--654, 2016.

\bibitem{Sezer_2013_IEEECommunicationsMagazine_ArewereadyforSDN?Implementationchallengesforsoftware-definednetworks}
S.~Sezer, S.~Scott-Hayward, P.~Chouhan, B.~Fraser, D.~Lake, J.~Finnegan,
  N.~Viljoen, M.~Miller, and N.~Rao, ``Are we ready for {SDN}? implementation
  challenges for software-defined networks,'' \emph{{IEEE} Communications
  Magazine}, vol.~51, no.~7, pp. 36--43, Jul. 2013.

\bibitem{McBride_2013_OpenNetworkingFoundation-ONFSOLUTIONBRIEF_SDNsecurityconsiderationsinthedatacenter}
M.~McBride, M.~Cohn, S.~Deshpande, M.~Kaushik, M.~Mathews, and S.~Nathan, ``Sdn
  security considerations in the data center,'' \emph{Open Networking
  Foundation-ONF SOLUTION BRIEF}, pp. 15--16, 2013.

\bibitem{Chi_2014_AnAMIThreatDetectionMechanismBasedonSDNNetworks}
\BIBentryALTinterwordspacing
P.-W. Chi, C.-T. Kuo, H.-M. Ruan, S.-J. Chen, and C.-L. Lei, ``An ami threat
  detection mechanism based on sdn networks,'' in \emph{Eighth International
  Conference on Emerging Security Information, Systems and Technologies
  (SECUWARE 2014)}.\hskip 1em plus 0.5em minus 0.4em\relax IARIA, Nov. 2014.
  [Online]. Available:
  \url{https://www.thinkmind.org/download.php?articleid=securware_2014_9_30_30142}
\BIBentrySTDinterwordspacing

\bibitem{Barbette_2015_FastUserspacePacketProcessing}
T.~Barbette, C.~Soldani, and L.~Mathy, ``Fast userspace packet processing,'' in
  \emph{2015 {ACM}/{IEEE} Symposium on Architectures for Networking and
  Communications Systems ({ANCS})}, IEEE.\hskip 1em plus 0.5em minus
  0.4em\relax {IEEE}, May 2015, pp. 5--16.

\bibitem{Gallenmuller_2015_ComparisonofFrameworksforHigh-performancePacketIO}
S.~Gallenmuller, P.~Emmerich, F.~Wohlfart, D.~Raumer, and G.~Carle,
  ``Comparison of frameworks for high-performance packet {IO},'' in \emph{2015
  {ACM}/{IEEE} Symposium on Architectures for Networking and Communications
  Systems ({ANCS})}, IEEE Computer Society.\hskip 1em plus 0.5em minus
  0.4em\relax {IEEE}, May 2015, pp. 29--38.

\bibitem{Hoiland-Jorgensen_2018_TheeXpressDataPathFastProgrammablePacketProcessingintheOperatingSystemKernel}
T.~H{\o}iland-J{\o}rgensen, J.~D. Brouer, D.~Borkmann, J.~Fastabend,
  T.~Herbert, D.~Ahern, and D.~Miller, ``The express data path: Fast
  programmable packet processing in the operating system kernel,'' in
  \emph{Proceedings of the 14th International Conference on emerging Networking
  {EXperiments} and Technologies - {CoNEXT} {'}18}, ACM.\hskip 1em plus 0.5em
  minus 0.4em\relax {ACM} Press, 2018, pp. 54--66.

\bibitem{Lorenz_2017_IEEECommunicationsMagazine_AnSDN/NFV-EnabledEnterpriseNetworkArchitectureOfferingFine-GrainedSecurityPolicyEnforcement}
C.~Lorenz, D.~Hock, J.~Scherer, R.~Durner, W.~Kellerer, S.~Gebert, N.~Gray,
  T.~Zinner, and P.~Tran-Gia, ``An {SDN}/{NFV}-enabled enterprise network
  architecture offering fine-grained security policy enforcement,''
  \emph{{IEEE} Communications Magazine}, vol.~55, no.~3, pp. 217--223, Mar.
  2017.

\bibitem{Farris_2019_IEEECommunicationsSurveys&Tutorials_ASurveyonEmergingSDNandNFVSecurityMechanismsforIoTSystems}
I.~Farris, T.~Taleb, Y.~Khettab, and J.~Song, ``A survey on emerging {SDN} and
  {NFV} security mechanisms for {IoT} systems,'' \emph{{IEEE} Communications
  Surveys {\&} Tutorials}, vol.~21, no.~1, pp. 812--837, 2019.

\bibitem{Li_2017_J.InternetServ.Inf.Secur._Application-awareandDynamicSecurityFunctionChainingforMobileNetworks}
G.~Li, H.~Zhou, G.~Li, and B.~Feng, ``Application-aware and dynamic security
  function chaining for mobile networks,'' \emph{J. Internet Serv. Inf.
  Secur.}, vol.~7, pp. 21--34, 2017.

\bibitem{Li_2018_IEEEAccess_RuleAnomaly-FreeMechanismofSecurityFunctionChainingin5G}
G.~Li, H.~Zhou, B.~Feng, G.~Li, H.~Zhang, and T.~Hu, ``Rule anomaly-free
  mechanism of security function chaining in 5g,'' \emph{{IEEE} Access},
  vol.~6, pp. 13\,653--13\,662, 2018.

\bibitem{Luizelli_2015_PiecingTogethertheNFVProvisioningPuzzleEfficientPlacementandChainingofVirtualNetworkFunctions}
M.~C. Luizelli, L.~R. Bays, L.~S. Buriol, M.~P. Barcellos, and L.~P. Gaspary,
  ``Piecing together the {NFV} provisioning puzzle: Efficient placement and
  chaining of virtual network functions,'' in \emph{2015 {IFIP}/{IEEE}
  International Symposium on Integrated Network Management ({IM})}.\hskip 1em
  plus 0.5em minus 0.4em\relax {IEEE}, May 2015.

\bibitem{Shameli-Sendi_2019_IEEETransactionsonServicesComputing_EfficientProvisioningofSecurityServiceFunctionChainingUsingNetworkSecurityDefensePatterns}
A.~Shameli-Sendi, Y.~Jarraya, M.~Pourzandi, and M.~Cheriet, ``Efficient
  provisioning of security service function chaining using network security
  defense patterns,'' \emph{{IEEE} Transactions on Services Computing},
  vol.~12, no.~4, pp. 534--549, Jul. 2019.

\bibitem{Milenkoski_2016_SecurityPositionPaperNetworkFunctionVirtualization}
\BIBentryALTinterwordspacing
A.~Milenkoski, B.~Jaeger, K.~Raina, M.~Harris, S.~Chaudhry, S.~Chasiri,
  V.~David, and W.~Liu, ``{Security Position Paper: Network Function
  Virtualization},'' Mar. 2016, published by Cloud Security Alliance (CSA) -
  Virtualization Working Group. [Online]. Available:
  \url{https://cloudsecurityalliance.org/download/security-position-paper-network-function-virtualization/}
\BIBentrySTDinterwordspacing

\bibitem{Kistowski_2018_RunTimePredictionofPowerConsumptionforComponentDeployments}
J.~von Kistowski, M.~Deffner, and S.~Kounev, ``Run-time prediction of power
  consumption for component deployments,'' in \emph{2018 {IEEE} International
  Conference on Autonomic Computing ({ICAC})}.\hskip 1em plus 0.5em minus
  0.4em\relax {IEEE}, Sep. 2018.

\bibitem{Ifflaender_2018_AddressingShortcomingsofExistingDDoSProtectionSoftwareUsingSoftware-DefinedNetworking}
L.~Iffländer, S.~Geißler, J.~Walter, L.~Beierlieb, and S.~Kounev,
  ``{Addressing Shortcomings of Existing DDoS Protection Software Using
  Software-Defined Networking},'' in \emph{Proceedings of the 9th Symposium on
  Software Performance 2018 (SSP'18)}, 11 2018.

\bibitem{Kistowski_2018_TeaStoreAMicroServiceReferenceApplicationforBenchmarkingModelingandResourceManagementResearch}
J.~von Kistowski, S.~Eismann, N.~Schmitt, A.~Bauer, J.~Grohmann, and S.~Kounev,
  ``{TeaStore: A Micro-Service Reference Application for Benchmarking, Modeling
  and Resource Management Research},'' in \emph{Proceedings of the 26th IEEE
  International Symposium on the Modelling, Analysis, and Simulation of
  Computer and Telecommunication Systems}, ser. MASCOTS '18, Sep. 2018.

\end{thebibliography}

\end{document}